\documentstyle{mn}

\input{epsf}

\newcommand{\mpc}{\rm {h^{-1}Mpc }}
\newcommand{\etal}{{\it et al.\ }}

\newcommand{\omav}{\bar{\omega}}

\newcommand{\ltsima}{$\; \buildrel < \over \sim \;$}
\newcommand{\lsim}{\lower.5ex\hbox{\ltsima}}
\newcommand{\gtsima}{$\; \buildrel > \over \sim \;$}
\newcommand{\gsim}{\lower.5ex\hbox{\gtsima}}

\begin{document}

\title[Comparison of the APM and the EDSGC 
Surveys]{Comparison of the Large Scale Clustering in the APM and the EDSGC 
Galaxy  Surveys}

\author[I.Szapudi and E. Gazta\~{n}aga]{Istv\'an Szapudi$^{1}$
 E. Gazta\~{n}aga $^{2}$
%\altaffilmark{1}}
%\affil{Fermi National Accelerator Laboratory}
%\affil{Theoretical Astrophysics Group}
%\affil{Batavia, IL 60510}
%\author{Enrique Gazta\~ naga{2}}
%\affil{Institut d'Estudis Espacials de Catalunya}
%\affil{Research Unit (CSIC)}
%\affil{Edf. Nexus-104 - c/ Gran Capit\'an 2-4}
%\affil{08034 Barcelona, Spain}
%\altaffiltext{1}{E-mail:\ szapudi@astro1.fnal.gov}
%\altaffiltext{2}{E-mail:\ gaztanaga@ieec.fcr.es}
\\
1. University of Durham, Department of Physics 
 South Road,  Durham DH1 3LE, United Kingdom               
\\
2. Institut d'Estudis Espacials de Catalunya, 	Research Unit (CSIC),
Edf. Nexus-104 - c/ Gran Capitan 2-4, 08034 Barcelona
}

\maketitle 
 
\def\mpc {h^{-1} {\rm Mpc}}
\def\impc {h {\rm Mpc}^{-1}}
\def\and  {{\it {et al.} }}
\def\rmd {{\rm d}}

\begin{abstract}

Clustering statistics are compared
in the Automatic Plate Machine (APM)
and the Edinburgh/Durham Southern Galaxy Catalogue (EDSGC)
angular galaxy surveys. 
Both surveys were independently constructed from
scans of the same adjacent UK IIIa--J Schmidt
photographic plates with the APM and COSMOS microdensitometers,
respectively. The comparison of these catalogs
is a rare practical opportunity to
study systematic errors, which cannot be achieved via
simulations or theoretical methods.
On intermediate scales, $0.1^\circ < \theta < 0.5^\circ$, 
we find good agreement for the cumulants or reduced moments of
counts in cells up to sixth order.
On larger scales there is a small disagreement 
due to edge effects in the EDSGC, which covers a smaller area.
On smaller scales, we find a 
significant disagreement that can only be attributed to differences
in the construction of the surveys, most likely the
dissimilar deblending of crowded fields. 
The overall agreement of the APM and EDSGC is encouraging,
and shows that the results for intermediate scales should be
fairly robust. On the other hand, the systematic deviations
found at small scales are
significant in a regime, where
comparison with theory and simulations is possible.
This is an important fact to bear in mind when planning the
construction of future digitized galaxy catalogs.

\end{abstract}


\section{Introduction}

Clustering measurements from galaxy catalogues have  become an important tool
to test models of structure formation. Large
sophisticated data sets are currently under analysis
or construction. To interpret high precision measurements of clustering,
a detailed understanding of the  uncertainties is
required. Errors can arise from finite size and geometry of
the catalog, such as discreteness, edge, and finite volume
effects (``cosmic errors''), from the insufficient
sampling of the measurement technique itself (''measurement errors''),
and finally, ``systematic errors'' arise from
data reduction, object detection, magnitude uncertainties, etc.
Studying the first two classes is by no means simple, 
but theoretical methods (e.g., Szapudi \& Colombi 1996,
hereafter SC96) and $N$-body simulations
yield reasonable estimates. Systematic errors are even more
difficult to investigate, and a unique opportunity is provided,
when the same raw data are reduced independently by two
research teams. The goal of this {\em Letter} is 
to seize on such an opportunity:
the APM and the EDSGC galaxy surveys were constructed independently from 
the same underlying photographic plates. In particular, we investigate
the degree of reproducibility of the higher order clustering measurements,
i.e. to what extent different choices during the construction 
of a galaxy catalog can lead to different estimates
of clustering.

The most wide spread tools to study clustering in a galaxy catalog
are the two-point correlation function, $\xi_2$, and the amplitudes
of the higher order correlation functions. These latter
are usually expressed in the form of hierarchical ratios:
$S_J= \xi_J/\xi_2^{J-1}$, where $\xi_J$ is the $J$-order correlation
function or reduced cumulant. 
The predictions for $S_J$'s in both perturbation theory and
$N$-body simulations 
\cite{peebles80,bern92,jbc93,bern94,gb95,bge95,cbh96,bg96,sqsl97}
can be used to test the gravitational instability picture,
 the form of the initial
conditions and the biasing parameters \cite{fg94,gf94}.
The $S_J$'s are more difficult to measure and interpret than
the two-point function, however, 
at low orders, 
they are less affected by intrinsic observational uncertainties,
like time evolution or projection effects.

In section \S2 we summarize the properties of the two catalogues,
the method of analysis and the actual comparison follows in
sections \S3 and \S4. \S5 discusses the implications of the results. 

\section{The APM and  Edinburgh/Durham Southern Galaxy Catalogues}
\label{sec:catalogues}

The APM Galaxy Survey covers 4300 square degrees on the sky 
and contains over 2 million galaxies to a limiting apparent 
magnitude of $b_{J} \le 20.5$ \cite{mad90a,mad90b,mad90c,mad96}.
It was constructed from
APM (a microdensitometer) scans of 188
adjacent UK IIIa--J Schmidt
photographic plates and reaches a limiting magnitude of $b_j=20.5$.
In an extensive analysis of the systematic errors involved in 
plate matching, Maddox {\it et al} (1996) have placed an upper limit 
of $\delta w(\theta) \sim 1 \times 10^{-3}$ on the likely 
contribution of the systematic errors to the angular 
correlations. The shape of the angular correlation function measured 
from the survey at scales of $\theta > 1^{\circ}$  
indicates that the universe contains more structure on 
large scales than is predicted by the standard Cold Dark 
Matter scenario (Maddox {\it et al} 1990c). The higher
order correlations in the APM were measured by 
\cite{gaz94,sdes95,ss97a}.

The EDSGC is a catalogue
of 1.5 million galaxies covering $\simeq1000$ square degrees centered on
the South Galactic Pole. The database was constructed from
COSMOS scans (a microdensitometer) of 60 adjacent UK IIIa--J Schmidt
photographic plates (a subset of the APM plates)
and also reaches a limiting magnitude of $b_{J,EDSGC}=20.5$.

The entire catalogue has $<10\%$ stellar contamination and is $\gsim95\%$
complete for galaxies brighter than $b_j=19.5$ \cite{heydon89}.
The two--point galaxy angular
correlation function measured from the EDSGC has been presented by
Collins, Nichol, \& Lumsden (1992) and Nichol \& Collins (1994).
The higher order correlations in the EDSGC were measured
by Szapudi, Meiksin, \& Nichol 1996, hereafter SMN96.

We emphasize that the raw data for both
catalogs comprise of the same UK IIIa-J Schmidt Plates
(a smaller subset in case of the EDSGC), while the hardware
to digitize the plates and the
the software to classify and detect objects, measure their apparent
magnitudes were different. 
In particular,  different methods of calibration,
plate-matching, deblending algorithms were employed.
As a consequence, there
is a small offset in the magnitude scales of the two catalogues 
\cite{nichol92}, even though a simple one-to-one mapping
can be established.

Magnitude cuts for the comparison
of the statistics were determined by practical considerations.
For the APM we follow G94 and use $ m_{\rm APM}=17-20$, which is  
half a magnitude brighter than the completeness limit.
For the EDSGC catalogue, which
is complete to about $ m_{\rm EDS}=20.3$ magnitude,
we follow SMN96 to use a magnitude cut of
$16.98 \le m_{\rm EDS} \le 19.8$,
which is again half a magnitude
brighter than the completeness limit.
Based on matching the surface densities 
listed in SDES, these
magnitude ranges approximately correspond to each other. This
facilitates the direct cross-comparison of the results.

\section{The Method of Analysis}

The calculation of the higher order correlation functions 
followed closely the method outlined in 
\cite{smn96}. It
consists of estimating the probability distribution of counts in cells,
calculation of the factorial moments, and extraction of the
normalized, averaged amplitudes of the $J$-point correlation functions.
For the most crucial first step the infinitely oversampling
algorithm of \cite{s97} was used. Only few of the most important
definitions are presented below.

The average of the $J$-point angular correlation functions on a scale $\ell$
is defined by
\begin{equation}
  \omav_J (\ell)=A(\ell)^{-J}\int d^2r_1\ldots d^2r_J 
~\omega_J(r_1,\ldots,r_J),
\end{equation}
where $\omega_J$ is the $J$-point correlation function in the two dimensional
survey, and $A(\ell)$ is the area of a square cell of size $\ell$.
The hierarchical ratios, $s_J$, are defined in the usual way,
\begin{equation}
   s_J = \frac{\omav_J}{\omav_2^{J-1}}.
\end{equation}
The raw counts in cells measurements are reduced
to a set consisting of $n,\omav_2, s_J$, 
which forms a suitable basis for subsequent
comparison of the statistics; $n$ denotes the average count in a cell.

Counts in cells were measured in square cells with sizes in
the range $0.015125^\circ-2^\circ$
(corresponding to $0.1-14\mpc$ with $D \simeq 400\mpc$,
the approximate depth of the catalogues). Practical
considerations determined this scale range: the upper scale
was chosen to minimize the edge effects from cut-out holes,
while the smallest scale approaches that of galaxy halos for the typical
depth of the catalogs. For details see \cite{smn96}.
Note that physical coordinates
were used in both surveys to eliminate the effects of distortion.

\section{Comparison}

The amplitudes of the measured $J$-point correlation
functions for $2 \ge J \ge 6$ are displayed on a series of
figures. To facilitate comparison with perturbation theory, 
angular scales in all graphs  were converted to an equivalent
circular cell size, $\theta$, i.e. $\pi \theta^2 = \ell^2$.
Note that square cells were used for the measurements, up to a small
deformation due to projection. This has a negligible 
effect through slightly differing form factors, which
cancels out anyway when comparing the  results from the two catalogs
with {\em each other}. The cell size in the APM pixel maps
is defined by dividing the full APM area over the number of cells.
The corresponding scale is about $5\%$ smaller than previously used in
G94 and SDES, where the cell size was defined 
as the mean equal area projection size.

The mean density of the EDSGC counts is about $10\%$ smaller than that of
the APM (see also SMN97). This is partially due to star mergers
which account to $5\%$ of the APM images in the $b_J=17-20$ slice
(Maddox \etal 1990). The remaining $5\%$ can be attributed to a small
difference in the depths due to a slight offset in the magnitude slices.

\begin{figure}
\centering
\centerline{\epsfysize=9.truecm
\epsfbox{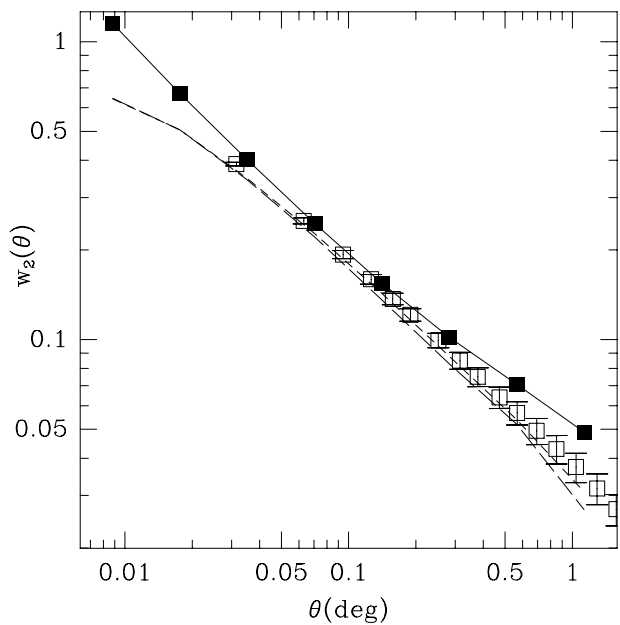}}
\caption[junk]{
The solid squares joined by continuous lines show our measurement of the 
angular $w_2$ over the EDSGC survey area with infinite sampling.
Open squares with errorbars display the
mean and variance of the $w_2$ measurements in four equal parts of the
APM survey, estimated with low sampling.
The short-dashed line corresponds to similar APM measurements
with infinite sampling.
The long-dashed line shows the APM results restricted to the 
EDSGC region measured with infinite sampling.}
\label{sigma}
\end{figure}

Figure \ref{sigma} shows the variance of counts-in-cells as 
a function of the cell radius in degrees. The full squares linked
by the solid line correspond to
the measurement in the EDSGC catalogue. The
small differences in the mean depth mentioned 
above should produce an upward shift of about
$10\%$ in the EDSGC correlation amplitude,
 which is confirmed by the Figure.
The open squares display the 
measurements by G94 for the full APM catalogue, while the short-dashed
line is the recalculation of the same with infinite sampling.
The long-dashed line is the measurement of a subregion of the
APM which overlaps with the EDSGC ($EDSGC\cap APM$).
The latter agrees well within the errors
with the full APM measurements and is slightly 
lower than the corresponding $w_2$ in the
EDSGC catalogue, roughly as expected from the mentioned differences.
There is an overall agreement between all estimates, at least on
large scales. On smaller scales the APM appears to produce slightly
lower values; this is probably related to the larger discrepancy
of the hierarchical amplitudes which will be discussed next.

\begin{figure}
\centering
\centerline{\epsfysize=9.truecm
\epsfbox{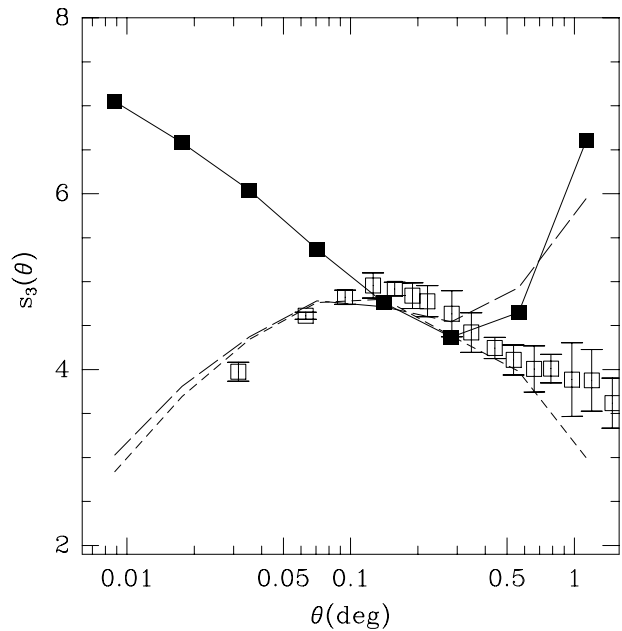}}
\caption[junk]{Same as in Figure \ref{sigma} for the 
the hierarchical skewness $s_3=w_3/w_2^2$.
The misalignment of the open and solid  squares at scales $gt 0.5$ degrees
is the  result of edge effects, as both correspond to smaller surveys.}
\label{s3}
\end{figure}

\begin{figure}
\centering
\centerline{\epsfysize=9.truecm
\epsfbox{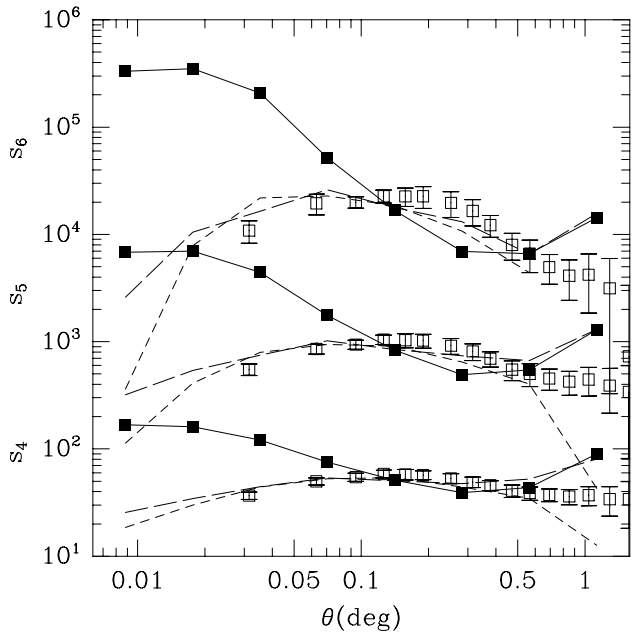}}
\caption[junk]{Same as in Figure \ref{s3} for $s_4$, $s_ 5$ and $s_5$.}
\label{sj}
\end{figure}

Figures \ref{s3}-\ref{sj} compare
the skewness,
$s_3$, and the higher order $s_J$'s, $J=4,5,6$.
The following discussion is equally applicable to all orders;
the separate graph for $J=3$ shows more details. Contrary to what happened
for $w_2$, a small difference in the depth
should not change the hierarchical ratios, as the depth cancels out in the
normalization (see \cite{gp77}).
The Figures follow this expectation. For scales  of
about $0.2^\circ$ to $2^\circ$ the agreement is good between the
full EDSGC and the same region of the APM ($EDSGC\cap APM$ region). 
The increase of the $s_J$'s at the largest scales ($\theta > 0.5^\circ$)
in the $EDSGC\cap APM$ region
is due to edge and finite volume effects:
a similar trend appears in the same region for both catalogues.
On these large scales, 
the full APM measurements are more accurate since
its larger area decreases cosmic 
errors.
Note that for the measurement represented with the short dashes
the edges of the catalog
were cut out generously to eliminate any possible
inhomogeneity. In addition, the masks was fully excluded,
while the original measurement followed a somewhat different procedure
(see G94 for details).
This could account for the slight difference at the largest scales.

The $S_J$'s measured in the $EDSGC\cap APM$ region of
the APM are compatible with the errors
of the full APM measurements at most scales.
At scales larger than $0.5^\circ$, edge effects start
to dominate the errors of the smaller sample.
For $0.1^\circ \ge\theta\ge 0.5^\circ$ the $EDSGC\cap APM$
region appears to produce slightly lower hierarchical ratios than
the full APM. These values in some cases are
outside of the formal errorbars. The reason for this is that dividing 
the sample into subsamples is an approximate estimate of the
errors, and it can lead to underestimation as the subsamples are
not fully independent. Moreover, for a non-Gaussian error distribution
%values outside the formal second order errorbar are not
values outside the formal errorbar are less
unlikely \cite{sc96}. 

At the smaller scales there is a significant
statistical difference between the APM and the  EDSGC. This is not 
due to finite volume effects, since it persists 
when only the same region of the sky is used. The identical
geometry with the same magnitude cut excludes
edge or discreteness effect as well, thus all cosmic errors.
The difference is not due to the method of estimation either,
since the original low sampling measurement by G94
gives similar results to the recalculation with infinite oversampling,
which fully eliminates measurement errors \cite{sc96,s97}. 
The only remaining possibility is that the results
should be attributed to systematics. 

\section{Discussion}

According to SMN97, insufficient sampling
can cause severe underestimation or the higher order $S_J$'s.
This could be a possible cause for 
the disagreement between  the EDGSC and the APM on
scales smaller than 0.2 degrees, 
since the original APM measurements by G94 were 
performed on density pixel map with resolution
given by the lowest scale shown at the figure. 
However, the infinite sampling $S_J$'s
are in good agreement with the original analysis by G94.
Although, as expected, the infinite oversampling results 
at small scales seem slightly
higher than the corresponding low sampling ones,
the Figures prove that this effect is not significant
and it can be discounted as the main reason for the disagreement
between the APM and the EDSGC.
The discrepancies on small scales are therefore due to 
intrinsic differences in the catalogues.
Since both catalogs use {\em same} raw photographic plates,
the difference discovered with the {\em same} statistical methods
must  lie with the different choices of hardware and software
during the  scans and/or the data reduction.
%The deblending algorithms is a good candidate to account for the detected 
%statistical difference, however,  this point needs
%further investigation.
The dissimilarity in the deblending algorithms is a particularly
good candidate to account for the detected statistical difference
(G. Efstathiou, private communication). However,  this point needs
further investigation.

Previous results and their interpretations on large scales seem
unaffected by the detected discrepancies.
In particular, both the APM and the EDSGC
higher order correlations are in  general agreement 
with perturbation theory
(G94, GF94, BGE95, SMN97). In summary,  the results support
qualitatively  scenarios with gravitational instability
arising from Gaussian initial conditions, with little
or no biasing.   Note that the EDSGC barely probes quasi-linear scales
($R> 8 \mpc$ or $\theta > 1^\circ$), thus  extended
perturbation theory, and results from 
$N$-body simulations have to be invoked
as a theoretical basis for comparison at smaller scales.
There is hint that, at least qualitatively, 
the EDSGC results at the smallest scales follow $N$-body simulations 
more closely, while the drop experienced in the APM reduced moments at
the same scales is unexpected, and could be an artificial effect.
The new generation of CCD based red-shift
and angular surveys, such as the SDSS, and 2DF, should
be able to clarify this situation and put tighter constraints
on biasing models.

\bigskip

{\bf Acknowledgments}

I.S. was supported by DOE and NASA through grant
NAG-5-2788 at Fermilab and by the PPARC rolling grant for 
Extragalactic Astronomy and Cosmology at Durham.
E.G. acknowledges support from 
supported by CSIC, DGICYT (Spain), project
PB93-0035, and CIRIT, grant GR94-8001 and
1996BEAI300192. We would like to thank both the APM
and EDSGC team for generously allowing us to use their
respective catalogs.

\def\apj { ApJ, }
\def\aap {A \& A, }
\def\ajs{ ApJS, }
\def\apjs{ ApJS, }
\def\mnras { MNRAS, }
\def\apjl { Ap. J. Let., }

\end{document}